\documentclass[aps,prl,preprint,groupedaddress,showpacs,amssymb,amsmath]{revtex4-1}

\newtheorem{theorem}{Theorem}

\newtheorem{proposition}[theorem]{Proposition}

\newtheorem{remark}{Remark}
\newtheorem{example}{Example}

\begin{document}


\title{$PT$-symmetric quantum mechanics is \emph{a} Hermitian quantum mechanics}


\author{Sungwook Lee}
\email[E-mail:]{sunglee@usm.edu}
\homepage[Home Page:]{http://www.math.usm.edu/lee/}
\affiliation{Department of Mathematics, University of Southern Mississippi}


\date{\today}

\begin{abstract}
The author discusses a different kind of Hermitian quantum mechanics, called $J$-Hermitian quantum mechanics. He shows that $PT$-symmetric quantum mechanics is indeed $J$-Hermitian quantum mechanics, and that time evolution (in the Krein space of states) is unitary if and only if Hamiltonian is $J$-Hermitian (or equivalently $PT$-symmetric). An issue with unitarity comes up when time evolution is considered in the Hilbert space of states rather than in the Krein space of states. The author offers possible scenarios with this issue.
\end{abstract}


\maketitle

\section{Prologue. $PT$-Symmetric Quantum Mechanics}

In standard quantum mechanics, Hamiltonian operators are required to be Hermitian. One reason is that the eigenvalues of a Hamiltonian operator are energies and they have to be real. Hermitian operators are indeed guaranteed to have all real eigenvalues. However, Hermitian operators may not be the only kind of operators whose eigenvalues are all real. There may be a kind of non-Hermitian Hamiltonian operators which are also guaranteed to have all real eigenvalues. If so, one may study a viable quantum theory with them. Carl M. Bender observed that non-Hermitian Hamiltonian operators such as
\begin{equation}
\label{eq:pt-hamiltonian}
 \hat H=-\frac{\hbar^2}{2m}\frac{\partial^2}{\partial x^2}+x^2(ix)^\epsilon,
\end{equation}
where $\epsilon$ is a non-negative real number, have all real eigenvalues \cite{Bender},\cite{Bender-Boettcher-Meisinger}. Such Hamiltonian operators are characterised as $PT$-symmetric Hamiltonians. Here $P$ and $T$ stand for the operators defined by
\begin{align*}
 P:\ x&\mapsto -x;\\
 T:\ i&\mapsto -i.
\end{align*}
$P$ is the parity and Carl M. Bender calls $T$ the time-reversal operator. $T$ may be viewed as the time-reversal operator since $i\mapsto -i$ has the same effect on the energy operator $\hat E=i\hbar\frac{\partial}{\partial t}$ as $t\mapsto -t$. A Hamiltonian $\hat H$ is said to be $PT$-symmetric if it is invariant under the composite operator $PT$. Note that the commutation relation
\begin{equation}
 [\hat x,\hat p]=i\hbar
\end{equation}
is $PT$-symmetric. While $PT$-symmetric quantum mechanics has been studied for a long time by many physicists, their studies mostly focused on the numerical investigation of particular $PT$-symmetric Hamiltonians \eqref{eq:pt-hamiltonian} and it still lacks mathematical rigor and general theory.

It turns out that $PT$-symmetric quantum mechanics is \emph{a} Hermitian quantum mechanics, but a different kind from the standard Hermitian quantum mechanics that we know. The author calls it $J$-Hermitian quantum mechanics. Here, $J$ is an involution on the space of states. If $J=I$, the identity transformation, the resulting quantum mechanics is the standard Hermitian quantum mechanics. In $J$-Hermitian quantum mechanics, the adjoint operator of a linear operator is defined analogously to that in the standard Hermitian quantum mechanics, and so the notion of self-adjointness or Hermiticity may be considered. The author shows that in $J$-Hermitian quantum mechanics time evolution is unitary if and only if Hamiltonian is self-adjoint ($J$-Hermitian). The author also shows that a Hamiltonian is $J$-Hermitian if and only if it is $PT$-symmetric. Hence, the eigenvalues of a $PT$-symmetric Hamiltonian are all real, and time evolution determined by a $PT$-symmetric Hamiltonian is unitary.
\section{$J$-Hermitian Quantum Mechanics and Krein Space}
For mathematical simplicity, we consider only $1$-dimensional quantum mechanics throughout this paper. Any part of our arguments do not depend on the spacial dimension. Please see the appendix \emph{Real Structure and Fundamental Symmetry} at the end of this paper for more mathematical details of rather heuristic discussion in this section.

Let us denote by $\mathcal{K}$ the complex vector space spanned by the eigenstates of a quantum system determined by Schr\"{o}dinger equation
\begin{equation}
\label{eq:schrodinger}
 i\hbar\frac{\partial\psi}{\partial t}=\hat H\psi.
\end{equation}
We assume that $\mathcal{K}$ is separable. Let $J:\mathcal{K}\longrightarrow\mathcal{K}$ be an involution i.e. $J^2=I$. Define an inner product $\langle\ ,\ \rangle$ on $\mathcal{K}$ as follows: For any $\varphi,\psi\in\mathcal{K}$,
\begin{equation}
\label{eq:innerproduct}
 \begin{aligned}
  \langle\varphi,\psi\rangle&:=\langle\varphi|J|\psi\rangle\\
  &=\int_{-\infty}^\infty\bar\varphi J\psi dx.
 \end{aligned}
\end{equation}
If $J=I$ then $\langle\varphi,\psi\rangle=\langle\varphi|\psi\rangle$, the usual Dirac braket in the standard quantum mechanics. In this paper, we consider $J=P$, the parity. That is, $J$ acts on $\psi(x,t)$ as
\begin{equation}
\label{eq:parity}
 J\psi(x,t)=\psi(-x,t).
\end{equation}
Then the inner product is written
\begin{equation}
\label{eq:pt-product}
 \langle\varphi,\psi\rangle=\int_{-\infty}^\infty\bar\varphi(x)\psi(-x)dx.
\end{equation}
Equivalently, by substitution $x\to -x$, we also have 
\begin{equation}
\label{eq:pt-product2}
\begin{aligned}
 \langle\varphi,\psi\rangle&=\langle J\varphi|\psi\rangle\\
 &=\int_{-\infty}^\infty\bar\varphi(-x)\psi(x) dx.
\end{aligned}
 \end{equation}
The inner product in the form of \eqref{eq:pt-product2} was introduced by Carl M. Bender. He calls it the $PT$-inner product. The inner product is a Hermitian product, i.e. it satisfies

\noindent IP1. $\langle\varphi,\psi\rangle=\overline{\langle\psi,\varphi\rangle}$ for any $\varphi,\psi\in\mathcal{K}$.

\noindent IP2. $\langle\varphi,a\psi_1+b\psi_2\rangle=a\langle\varphi,\psi_1\rangle+b\langle\varphi,\psi_2\rangle$ for any $\varphi,\psi_1,\psi_2\in\mathcal{K}$ and $a,b\in\mathbb{C}$.

The squared norm $||\psi||^2=\int_{-\infty}^\infty\bar\psi(x)\psi(-x)dx$ can be positive, zero, or negative, so $\langle\ ,\ \rangle$ is an indefinite Hermitian product. The space $\mathcal{K}$ may be decomposed to
$$\mathcal{K}=\mathcal{K}^+\dotplus\mathcal{K}^-,$$
where $\mathcal{K}^+$ is spanned by the eigenstates $\psi^+_n$, $n=1,2,\cdots$ such that $\langle\psi^+_m,\psi^+_n\rangle=\delta_{mn}$ and $\mathcal{K}^-$ is spanned by the eigenstates $\psi^-_n$, $n=1,2,\cdots$ such that $\langle\psi^-_m,\psi^-_n\rangle=-\delta_{mn}$. Here, $\dotplus$ denotes orthogonal direct sum. Clearly $\mathcal{K}$ cannot be a (pre-)Hilbert space. It is called a \emph{Krein space} in mathematics literature \cite{Bognar}. Eigenstates of a $J$-Hermitian quantum mechanical system with vanishing squared norm cannot be physical for they are with complex eigenvalues. This will be proved in the next section. The involution $J$ is required to satisfy: for each $n=1,2,\cdots$,
\begin{equation}
\label{eq:fundsymm}
\begin{aligned}
 J\psi^+_n(x,t)&=\psi^+_n(-x,t)=\psi^+_n(x,t);\\
 J\psi^-_n(x,t)&=\psi^-_n(-x,t)=-\psi^-_n(x,t).
\end{aligned}
\end{equation}
The equations \eqref{eq:fundsymm} indicate that $\psi^+_n(x,t)$ are even functions with respect to $x$, while $\psi^-_n(x,t)$ are odd functions with respect to $x$. An involution on a Krein space satisfying \eqref{eq:fundsymm} is called the \emph{fundamental symmetry} in mathematics literature \cite{Bognar}. The fundamental symmetry $J$ is used to define a positive definite inner product $\langle\ ,\ \rangle_J$ on the Krein space $\mathcal{K}$: for any $\varphi,\psi\in\mathcal{K}$,
\begin{equation}
\label{eq:j-product}
\langle\varphi,\psi\rangle_J:=\langle\varphi,J\psi\rangle=\langle J\varphi,\psi\rangle=\langle\varphi|\psi\rangle.
\end{equation}
This inner product \eqref{eq:j-product} is called the $J$-inner product \cite{Bognar}. This $J$-inner product allows us to avoid awkward negative probability that was resulted by the indefinite Hermitian product \eqref{eq:pt-product}. $\mathcal{K}$ together with the $J$-inner product becomes a separable pre-Hilbert space and hence we may consider the Hilbert space of states as its completion. The $J$-inner product is the same as Dirac braket, so one may think that $J$-Hermitian quantum mechanics would be the same as the standard Hermitian quantum mechanics after all. However, it may not be the case. There are indeed some differences. One difference is that in $J$-Hermitian quantum mechanics, eigenstates consist of the same number (cardinality $\aleph_0$) of even functions and odd functions, while there is no such requirement on eigenstates in the standard Hermitian quantum mechanics. Another difference is that, as we shall see later, in $J$-Hermitian quantum mechanics, Hermitian operators may be complex, 
while in 
the standard Hermitian quantum mechanics, Hermitian operators must be real. Geometrically, $J$-Hermitian quantum mechanics and standard Hermitian quantum mechanics exhibit very distinct symmetries. This will be discussed elsewhere.
\begin{example}
 As well known, the de Broglie wave functions
 \begin{equation}
 \psi_n(x,t)=\frac{1}{\sqrt{L}}e^{i\frac{2n\pi}{L}x}e^{-i\frac{E_n}{\hbar} t},\ -L\leq x\leq L,\ n=1,2,\cdots
 \end{equation}
may serve as eigenstates of a free particle in the standard Hermitian quantum mechanics. In $J$-Hermitian quantum mechanics, the eigenstates of a free particle may be given by
\begin{equation}
 \begin{aligned}
  \psi^+_n(x,t)&=\frac{1}{\sqrt{L}}e^{-i\frac{E^+_n}{\hbar} t}\cos\frac{2n\pi}{L}x;\\
  \psi^-_n(x,t)&=\frac{1}{\sqrt{L}}e^{-i\frac{E^-_n}{\hbar} t}\sin\frac{2n\pi}{L}x,
 \end{aligned}\label{}
\end{equation}
where $-L\leq x\leq L$, $E^+_n=E^-_n=\frac{\hbar^2}{2m}\frac{4n^2\pi^2}{L^2}$, $n=1,2,\cdots$. The $\psi^+_n$ and $\psi^-_n$ satisfy
\begin{align*}
 \langle\psi^+_m,\psi^+_n\rangle&=\delta_{mn},\\
 \langle\psi^-_m,\psi^-_n\rangle&=-\delta_{mn},\\
 \langle\psi^+_m,\psi^-_n\rangle&=0.
\end{align*}
For each $n=1,2,\cdots$, let $\psi^+_n(x)=\frac{1}{\sqrt{L}}\cos\frac{2n\pi}{L}x$ and $\psi^-_n(x)=\frac{1}{\sqrt{L}}\sin\frac{2n\pi}{L}x$. Let $\mathcal{K}^+$ and $\mathcal{K}^-$ be spaces spanned by the orthonormal bases $\{\psi^+_n(x):-L\leq x\leq L,\ n=1,2,\cdots\}$ and $\{\psi^-_n(x):-L\leq x\leq L,\ n=1,2,\cdots\}$, respectively. Then $\mathcal{K}=\mathcal{K}^+\dotplus\mathcal{K}^-$ is a Krein space.

With $J$-inner product \eqref{eq:j-product}, $\mathcal{K}$ becomes a real pre-Hilbert space. Although its completion is a real Hilbert space, both standard Hermitian quantum mechanics and Hermitian quantum mechanics describe the same physics of a free particle motion. It is indeed expected since the kinetic energy operator $\hat p^2=-\frac{\hbar^2}{2m}\frac{\partial^2}{\partial x^2}$ is both Hermitian and $J$-Hermitian operators. (We discuss $J$-Hermitian operators in the following section.)
\end{example}
\begin{example}
 The Hamiltonian for quantum harmonic oscillator
 \begin{equation}
  \hat H=-\frac{\hbar^2}{2m}\frac{\partial^2}{\partial x^2}+\frac{m}{2}\omega^2\hat x^2
 \end{equation}
is both Hermitian and $J$-Hermitian. The stationary Schr\"{o}dinger equation $\hat H\psi=E\psi$ determines the eigenstates
\begin{equation}
 \psi_n(\xi)=2^{-\frac{n}{2}}\pi^{-\frac{1}{4}}(n!)^{-\frac{1}{2}}e^{-\frac{\xi^2}{2}}H_n(\xi),
\end{equation}
where $\xi=\sqrt{\frac{m\omega}{\hbar}}x$, and eigenvalues (energies)
\begin{equation}
 E_n=\left(n+\frac{1}{2}\right)\hbar\omega
\end{equation}
for $n=0,1,2,\cdots$.
The Hermite polynomials $H_n(x)$ satisfy the parity relation
\begin{equation}
 H_n(x)=(-1)^nH_n(-x).
\end{equation}
Let $\psi^+_n(\xi)=\psi_{2n}(\xi)$ and $\psi^-_n(\xi)=\psi_{2n+1}(\xi)$ for $n=0,1,2,\cdots$. Then $\langle\psi^+_m,\psi^+_n\rangle=\delta_{mn}$, $\langle\psi^-_m,\psi^-_n\rangle=-\delta_{mn}$, $\langle\psi^+_m,\psi^-_n\rangle=0$. Let $\mathcal{K}^+$ and $\mathcal{K}^-$ be spaces spanned by the orthonormal bases $\{\psi^+_n(\xi):,n=0,1,2,\cdots\}$ and $\{\psi^-_n(\xi):n=0,1,2,\cdots\}$, respectively. Then $\mathcal{K}=\mathcal{K}^+\dotplus\mathcal{K}^-$ is a Krein space.

With $J$-inner product \eqref{eq:j-product}, the Hilbert space resulted from the completion of $\mathcal{K}$ is the same Hilbert space of states of quantum harmonic oscillator in the standard Hermitian quantum mechanics.
\end{example}

\section{Adjoint Operators and Self-Adjointness}
Let $A:\mathcal{K}\longrightarrow\mathcal{K}$ be a bounded linear operator on a Krein space $\mathcal{K}$. Define its \emph{adjoint} to be the operator $A^\ast:\mathcal{K}\longrightarrow\mathcal{K}$ that has the property
\begin{equation}
 \langle\varphi,A\psi\rangle=\langle A^\ast\varphi,\psi\rangle
\end{equation}
or
\begin{equation}
 \int_{-\infty}^\infty\bar\varphi J(A\psi)dx=\int_{-\infty}^\infty\overline{A^\ast\varphi}J\psi dx
\end{equation}

for all $\varphi,\psi\in\mathcal{K}$. The adjoint satisfies the properties:

\noindent A1. $(A+B)^\ast=A^\ast+B^\ast$,

\noindent A2. $(\lambda A)^\ast=\bar\lambda A^\ast$,

\noindent A3. $(AB)^\ast=B^\ast A^\ast$,

\noindent A4. $A^{\ast\ast}=A$,

\noindent A5. if $A$ is invertible, then $(A^{-1})^\ast=(A^\ast)^{-1}$.

A bounded linear operator $A: \mathcal{K}\longrightarrow\mathcal{K}$ is said to be \emph{self-adjoint} or \emph{$J$-Hermitian} or simply \emph{Hermitian} (in case there is no confusion with the notion of standard Hermitian operators) if $A=A^\ast$, i.e. for any $\varphi,\psi\in\mathcal{K}$
\begin{equation}
 \langle A\varphi,\psi\rangle=\langle \varphi,A\psi\rangle
\end{equation}
or
\begin{equation}
\label{eq:hermitian}
 \int_{-\infty}^\infty(\overline{A\varphi})J\psi dx=\int_{-\infty}^\infty\bar\varphi J(A\psi)dx.
\end{equation}

Denote by $\langle\hat L\rangle$ the average value of an observable $L$ represented by an operator $\hat L$ in a state $\psi$:
\begin{equation}
\begin{aligned}
 \langle\hat L\rangle&=\langle\psi,\hat L\psi\rangle\\
 &=\langle\psi|J|\hat L\psi\rangle.
 \end{aligned}
\end{equation}
Suppose that $\hat L$ is Hermitian. Then
\begin{align*}
\langle\hat L\rangle&=\int_{-\infty}^\infty\bar\psi J(\hat L\psi)dx\\
 &=\int_{-\infty}^\infty(\overline{\hat L\psi})J\psi dx\ \mbox{[by \eqref{eq:hermitian}]}\\
 &=\overline{\int_{-\infty}^\infty(\hat L\psi)J\bar\psi dx}\\
 &=\overline{\int_{-\infty}^\infty\bar\psi J(\hat L\psi)dx}\\
 &=\overline{\langle\hat L\rangle}
 \end{align*}
As in the standard Hermitian quantum mechanics, the average of a Hermitian operator is guaranteed to be real in $J$-Hermitian quantum mechanics. In particular, the eigenvalues of a Hermitian operator are guaranteed to be all real. Although it may seem redundant, let us prove the reality of eigenvalues of a Hermitian operator to make a point on the distinctive nature of $J$-Hermitian quantum mechanics. Let $\hat L$ be a Hermitian operator. Let $\psi$ be an eigenfunction of $\hat L$ with eigenvalue $\lambda$. Then
$$\hat L\psi=\lambda\psi.$$
\begin{align*}
 \langle\hat L\psi,\psi\rangle&=\langle\lambda\psi,\psi\rangle\\
&=\bar\lambda\langle\psi,\psi\rangle,
\end{align*}
and
\begin{align*}
 \langle\psi,\hat L\psi\rangle&=\langle\psi,\lambda\psi\rangle\\
&=\lambda\langle\psi,\psi\rangle.
\end{align*}
Since $\hat L$ is Hermitian,
$$\bar\lambda\langle\psi,\psi\rangle=\lambda\langle\psi,\psi\rangle.$$
In order for us to show that $\lambda=\bar\lambda$ i.e. $\lambda$ is real, we must additionally require that $||\psi||^2=\langle\psi,\psi\rangle\ne 0$. In fact, eigenstates of a Hermitian operator with vanishing squared norm are with complex eigenvalues as mentioned in the preceding section. To show this, let $\hat L$ be a Hermitian operator and $\lambda$ an eigenvalue of $\hat L$. Suppose that $\lambda$ is a non-real complex number $\lambda=a+ib$ where $a$ and $b\ne 0$ are real. Let $\psi$ be an eigenfunction of $\hat L$ with eigenvalue $\lambda$. Then
$$\hat L\psi=\lambda\psi.$$
Multiplying this equation by $\bar\lambda$, we obtain
$$\bar\lambda\hat H\psi=|\lambda|^2\psi$$
or
$$\frac{\bar\lambda}{|\lambda|^2}\hat H\psi=\psi.$$
\begin{align*}
 \langle\psi,\psi\rangle&=\left\langle\frac{\bar\lambda}{|\lambda|^2}\hat L\psi,\psi\right\rangle\\
 &=\frac{\lambda}{|\lambda|^2}\langle\hat L\psi,\psi\rangle\\
 &=\frac{\lambda}{|\lambda|^2}\langle\psi,\hat L\psi\rangle\ [\mbox{since $\hat L$ is Hermitian}]\\
 &=\frac{\lambda}{|\lambda|^2}\langle\psi,\lambda\psi\rangle\\
 &=\frac{\lambda^2}{|\lambda|^2}\langle\psi,\psi\rangle.
\end{align*}
This implies that
$$(\lambda^2-|\lambda|^2)\langle\psi,\psi\rangle=0.$$
Since $\lambda$ is a non-real complex number, $\lambda^2-|\lambda|^2\ne 0$ and hence $\langle\psi,\psi\rangle=0$.

\section{Time Evolution in $J$-Hermitian Quantum Mechanics}
Let $\hat H$ be a time-independent Hamiltonian. Denote by $\hat U(x,t)$ the time evolution operator of a system determined by Hamiltonian $\hat H$.Then
$$\psi(x,t)=\hat U(x,t)\psi(x,0).$$
$\hat U(x,t)$ is given by Schr\"{o}dinger equation \eqref{eq:schrodinger} as
\begin{equation}
\label{eq:propagator}
 \hat U(x,t)=\exp\left(-\frac{1}{\hbar}\hat H t\right).
\end{equation}
We require $\hat U(x,t)$ to be unitary:
\begin{align*}
 \langle\psi(x,t),\psi(x,t)\rangle&=\langle\hat U(x,t)\psi(x,0),\hat U(x,t)\psi(x,0)\rangle\\
 &=\langle\hat U^\ast(x,t)\hat U(x,t)\psi(x,0),\psi(x,0)\rangle\\
 &=\langle e^{\frac{i}{\hbar}\hat H^\ast t}e^{-\frac{i}{\hbar}\hat H t}\psi(x,0),\psi(x,0)\rangle\\
 &=\langle \psi(x,0),\psi(x,0)\rangle.
\end{align*}
\begin{theorem}[Theorem 4.2 in \cite{Bognar}]
\label{thm:unitary}
 Let $\mathcal{K}$ be a Krein space. Then $\hat U:\mathcal{K}\longrightarrow\mathcal{K}$ is unitary if and only if $\hat U$ is invertible and $\hat U^{-1}=\hat U^\ast$.
\end{theorem}
By Theorem \ref{thm:unitary}, $\hat U(x,t)$ is unitary if and only if 
\begin{equation}
\label{eq:unitary2}
\hat U^\ast (x,t)U(x,t)=e^{\frac{i}{\hbar}\hat H^\ast t}e^{-\frac{i}{\hbar}\hat H t}=I
\end{equation}
for all $t$.
If
\begin{equation}
\label{eq:j-hermitian}
 \hat H^\ast=\hat H
\end{equation}
i.e. $\hat H$ is Hermitian, then clearly time evolution operator $\hat U(x,t)$ \eqref{eq:propagator} is unitary. Conversely, if $\hat U(x,t)$ is unitary then by differentiating \eqref{eq:unitary2} with respect to $t$ at $t=0$, we obtain \eqref{eq:j-hermitian}. Therefore, in $J$-Hermitian quantum mechanics time evolution is unitary if and only if Hamiltonian is Hermitian.
\begin{remark}
 If $\hat U(x,t)$ is unitary, then $\mathcal{K}^+$ and $\mathcal{K}^-$ are $\hat U(x,t)$-invariant subspaces of the Krein space $\mathcal{K}$.
\end{remark}
\section{$J$-Hermiticity and $PT$-symmetry}
The author mentioned earlier some major differences between $J$-Hermitian quantum mechanics and the standard Hermitian quantum mechanics. Another difference is that the momentum and the position operators $\hat p=-i\hbar\frac{\partial}{\partial x}$ and $\hat x$ are not Hermitian in $J$-Hermitian quantum mechanics but are skew-Hermitian. The kinetic energy operator $\hat T=\frac{\hat p^2}{2m}=-\frac{\hbar^2}{2m}\frac{\partial^2}{\partial x^2}$ is still Hermitian. The momentum operator $\hat p$ is $PT$-symmetric while the position operator $\hat x$ is not. Clearly, the kinetic energy operator $\hat T$ is $PT$-symmetric. So, there is a difference between $J$-Hermitian quantum mechanics and $PT$-symmetric quantum mechanics. However, we now show that $PT$-symmetric quantum mechanics is essentially $J$-Hermitian quantum mechanics.

Let $\hat V(x)$ be a potential energy operator which acts on $\psi(x)$ by multiplication. For any $\psi_1(x),\psi_2(x)\in\mathcal{K}$,
\begin{align*}
 \langle\psi_1(x),\hat V(x)\psi_2(x)\rangle&=\int_{-\infty}^\infty\bar\psi_1(x)J(\hat V(x)\psi_2(x))dx\\
 &=\int_{-\infty}^\infty\bar\psi_1(x)\hat V(-x)\psi_2(-x)dx\\
 &=\int_{-\infty}^\infty \hat V(-x)\bar\psi_1(x)\psi_2(-x)dx\\
 &=\int_{-\infty}^\infty\overline{\overline{\hat V}(-x)\psi_1(x)}J\psi_2(x)dx\\
 &=\langle\overline{\hat V}(-x)\psi_1(x),\psi_2(x)\rangle.
\end{align*}
Hence, we obtain
\begin{equation}
\label{eq:pt-potential}
 \hat V^\ast(x)=\overline{\hat V}(-x)=PT[\hat V(x)].
\end{equation}
Therefore, $\hat V(x)$ is Hermitian if and only if $\hat V(x)$ is $PT$-symmetric. That is, Hamiltonian $\hat H$ of the form
\begin{equation}
\label{eq:hamiltonian}
 \hat H=-\frac{\hbar^2}{2m}\frac{\partial^2}{\partial x^2}+\hat V(x)
\end{equation}
is Hermitian if and only if it is $PT$-symmetric.
\begin{remark}
 In mathematics, any complex-valued function $f(x)$ satisfying
 \begin{equation}
  f(-x)=\bar f(x)
 \end{equation}
is called a \emph{Hermitian function}. It can be easily seen that the real part of a Hermitian function is an even function and the imaginary part of a Hermitian function is an odd function.
\end{remark}

$J$-Hermitian operators may be complex, while standard Hermitian operators must be real.
\begin{example}
 $\hat V(x)=ix^3$ is $J$-Hermitian (or $PT$-symmetric).
\end{example}

\section{The Conservation of Particle Number}
We showed earlier that time evolution is unitary if and only if Hamiltonian is Hermitian. Physically unitarity means that the sum of probabilities of all possible outcomes of the existence of a particle in a state $\psi$ does not change over time. This is equivalent to saying that particles are not created or annihilated i.e. the number of particles is conserved, so that if the probability density in a volume element changes in time, then a current flows through the surface of the volume element analogously to Gau\ss' law. In this section, we show more directly that particle number is conserved if and only if Hamiltonian is $PT$-symmetric (or equivalently Hermitian).

In $J$-Hermitian quantum mechanics, the probability density is given by
\begin{equation}
 w(x,t)=\bar\psi(x,t) J\psi(x,t)=\bar\psi(x,t)\psi(-x,t). 
\end{equation}
If we demand that no particles be created or annihilated, then we have a continuity equation hold.
\begin{equation}
\label{eq:continuity}
 \frac{\partial w}{\partial t}+\frac{\partial \jmath}{\partial x}=0.
\end{equation}
We deduce the \emph{particle current density} $\jmath$. From the time-dependent Schr\"{o}dinger equation \eqref{eq:schrodinger} with $\hat H$ in \eqref{eq:hamiltonian}, we have
\begin{align*}
 \frac{\partial w(x,t)}{\partial t}&=\frac{\partial\bar\psi(x,t)}{\partial t}\psi(-x,t)+\bar\psi(x,t)\frac{\partial\psi(-x,t)}{\partial t}\\
 &=\frac{i}{\hbar}[\overline{\hat H}\bar\psi(x,t)]\psi(-x,t)-\frac{i}{\hbar}\bar\psi(x,t)\hat H(-x)\psi(-x,t)\\
 &=-\frac{i\hbar}{2m}\frac{\partial}{\partial x}\left\{\frac{\partial\bar\psi(x,t)}{\partial x}\psi(-x,t)-\frac{\partial\psi(-x,t)}{\partial x}\bar\psi(x,t)\right\}+\\\frac{i}{\hbar}[\overline{\hat V}(x)-\hat V(-x)]\bar\psi(x,t)\psi(-x,t).
\end{align*}
 Let
 \begin{equation}
 \jmath:=\frac{i\hbar}{2m}\left\{\frac{\partial\bar\psi(x,t)}{\partial x}\psi(-x,t)-\frac{\partial\psi(-x,t)}{\partial x}\bar\psi(x,t)\right\}.
 \end{equation}
 Then in order for the continuity equation \eqref{eq:continuity} to hold, $\overline{\hat V}(x)=V(-x)$ or $\hat V(x)=\overline{\hat V}(-x)$, i.e. $V(x)$ is $PT$-symmetric (or equivalently Hermitian).
 \section{Time Evolution with $J$-Inner Product}
 In this section, we study time evolution of a state with $J$-inner product \eqref{eq:j-product}. So, we expect that the propagator of time evolution is a unitary operator on the Hilbert space of states rather than one on the Krein space of states.
 
 First, we need the following property for our discussion of time evolution.
 \begin{proposition}
  Let $A:\mathcal{K}\longrightarrow\mathcal{K}$ be a bounded linear operator. Then
  \begin{equation}
  \label{eq:adjoint}
   (JA)^\ast=JA^\ast,
  \end{equation}
where $J$ is the fundamental symmetry.
 \end{proposition}
\noindent{\it Proof.}
  For any $\varphi,\psi\in\mathcal{K}$,
  \begin{align*}
   \langle(JA)^\ast\varphi,\psi\rangle&=\langle\varphi,(JA)\psi\rangle\\
   &=\langle\varphi,J(AJ\psi)\rangle\\
   &=\langle J\varphi,AJ\psi\rangle\\
   &=\langle A^\ast J\varphi,J\psi\rangle\\
   &=\langle J(A^\ast J\varphi),\psi\rangle\\
   &=\langle JA^\ast\varphi,\psi\rangle.
  \end{align*}
Hence, $$(JA)^\ast=JA^\ast.$$

Let $\psi(x,t)=\hat U(x,t)\psi(x,0)$ where $\hat U(x,t)$ is the time evolution operator \eqref{eq:propagator}. Let us require that time evolution is unitary with respect to $J$-inner product \eqref{eq:j-product}:
\begin{align*}
 \langle\psi(x,t),J\psi(x,t)\rangle&=\langle\hat U(x,t)\psi(x,0),J[\hat U(x,t)\psi(x,0)]\rangle\\
&=\langle\hat U(x,t)\psi(x,0),[J\hat U(x,t)]J\psi(x,0)\rangle\\
&=\langle[J\hat U(x,t)]^\ast\hat U(x,t)\psi(x,0),J\psi(x,0)\rangle\\
&=\langle[J\hat U^\ast(x,t)]\hat U(x,t)\psi(x,0),J\psi(x,0)\rangle\ \mbox{[by \eqref{eq:adjoint}]}\\
&=\langle e^{\frac{i}{\hbar}\hat H^\ast(-x)t}e^{-\frac{i}{\hbar}\hat H(x)t}\psi(x,0),J\psi(x,0)\rangle\\
&=\langle\psi(x,0),J\psi(x,0)\rangle
\end{align*}
if and only if
\begin{equation}
\label{eq:j-unitary}
e^{\frac{i}{\hbar}\hat H^\ast(-x)t}e^{-\frac{i}{\hbar}\hat H(x)t}=I
\end{equation}
for all $t$. Here the notation $\hat H(x)$ means the dependence of $\hat H$ on $x$ and it does not mean the action of $\hat H$. If
\begin{equation}
\label{eq:j-unitary2}
\hat H^\ast(-x)=\hat H(x) 
\end{equation}
then clearly time evolution operator $\hat U(x,t)$ \eqref{eq:propagator} is unitary with respect to $J$-inner product. Conversely, if $\hat U(x,t)$ is unitary with respect to $J$-inner product, then by differentiating \eqref{eq:j-unitary} with respect to $t$ at $t=0$, we obtain \eqref{eq:j-unitary2}. Therefore, time evolution is unitary with respect to $J$-inner product if and only if Hamiltonian satisfies \eqref{eq:j-unitary2}. \eqref{eq:j-unitary2} may be written as
\begin{equation}
 J\hat H^\ast=\hat H.
\end{equation}

Suppose that $\hat H$ is a Hamiltonian as in \eqref{eq:hamiltonian} and that it satisfies \eqref{eq:j-unitary2}. Kinetic energy operator $\hat T(x)=-\frac{\hbar^2}{2m}\frac{\partial^2}{\partial x^2}$ is $J$-Hermitian and it is also invariant under parity, so we see that $\hat T(x)$ satisfies
$$\hat T^\ast(-x)=\hat T(x).$$
Hence, potential energy operator $\hat V(x)$ must satisfy
$$\hat V^\ast(-x)=\hat V(x).$$
Together with \eqref{eq:pt-potential}, we obtain
$$\overline{\hat V}(x)=\hat V(x),$$
i.e. the potential energy operator $\hat V(x)$ must be real. This means that complex Hamiltonians such as
$$\hat H=-\frac{\hbar^2}{2m}\frac{\partial^2}{\partial x^2}+ix^3$$
would violate unitarity if one considers time evolution with $J$-inner product. Could this be a sign of trouble for $J$-Hermitian or PT-symmetric quantum mechanics? The answer could be both yes and no. The author thinks that it depends on one's perspective. There may be possibly two different perspectives on this issue:

\noindent {\bf Perspective 1.} In $PT$-symmetric quantum mechanics, negative probabilities are merely an artifact of the indefinite Hermitian inner product \eqref{eq:pt-product} and  no physically meaningful notion for negative probabilities is offered. So one must use $J$-inner product for doing physics with PT-symmetric quantum mechanics. Therefore, time evolution in the Hilbert space of states rather than in the Krein space of states should be considered to be physical.

\noindent {\bf Consequence:} In this perspective, one should only allow $J$-Hermitian Hamiltonians that are real. Thus, the resulting quantum mechanics is essentially the same as the standard Hermitian quantum mechanics. $J$-Hermitian or equivalently $PT$-symmetric quantum mechanics offers no new physics with no apparent advantages over the standard Hermitian quantum mechanics.

\noindent {\bf Perspective 2.} Krein space structure is intrinsic to $J$-Hermitian or $PT$-symmetric quantum mechanics, so time evolution only needs to obey the symmetry of the Krein space of states, i.e. its propagators are not required to be unitary operators on the Hilbert space of states but are required to be unitary operators on the Krein space of states. One uses $J$-inner product only for probabilistic interpretations when needed.

\noindent {\bf Consequence:} In this perspective, one may allow complex $J$-Hermitian Hamiltonians which are not permitted in the standard Hermitian quantum mechanics. (In the standard Hermitian quantum mechanics, time evolution determined by a complex Hamiltonian is not unitary. It is also well-known in the standard Hermitian quantum mechanics that for particle number to be conserved the potential energy operator $\hat V(x)$ must be real.) $J$-Hermitian quantum mechanics offers a whole new class of physically and mathematically consistent Hamiltonians.

 \section{Epilogue}
The author discussed $J$-Hermitian quantum mechanics which exhibits distinctive mathematical and physical nature. The notion of self-adjointness or Hermiticity may be considered in $J$-Hermitian quantum mechanics analogously to the standard Hermitian quantum mechanics. As in the standard Hermitian quantum mechanics, time evolution is unitary if and only if Hamiltonian is Hermitian in $J$-Hermitian quantum mechanics. It turns out that $PT$-symmetric quantum mechanics is essentially $J$-Hermitian quantum mechanics, thereby the unitarity of $PT$-symmetric quantum mechanics is also proved. $PT$-symmetric quantum mechanics appears to be a consistent alternative quantum theory, and $J$-Hermitian quantum mechanics may serve as mathematically rigorous general theory for $PT$-symmetric quantum mechanics. However, $PT$-symmetric quantum mechanics may not be completely out of the woods yet. An issue with unitarity comes up when time evolution determined by a $PT$-symmetric Hamiltonian is considered with $J$-inner 
product in which case its propagator is a unitary operator on the Hilbert space of states rather than one on the Krein space of states. It turns out that in order for time evolution to be unitary with respect to $J$-inner product the $PT$-symmetric Hamiltonian must be real, in which case $PT$-symmetric quantum mechanics is essentially the same as the standard Hermitian quantum mechanics. The biggest hurdle for $PT$-symmetric quantum mechanics is whether complex Hamiltonians may be allowed. In the preceding section, the author offered two possible scenarios with corresponding physical consequences. The choice between the two scenarios depends on one's perspective of what a physical quantum theory should be. The author personally does not think complex potential energy is physical. But who knows? Maybe J. B. S. Haldane (5 November 1892 - 1 December 1964) was right to say \emph{``The universe is not only stranger than we imagine, it is stranger than we can imagine.''} After all, the final verdict is not ours 
but the 
mother nature's to make. 

After the author posted the first version of this paper on arXiv (\url{arXiv:1312.7738}), it was brought to his attention that Toshiaki Tanaka\cite{Tanaka} has done an extensive study on the relationship between $PT$-symmetry and $J$-Hermiticity (he calls it $P$-Hermiticity), and some of its consequences that the author discussed in this paper including the reality of eigenvalues and unitarity. However, his study was restricted on the Krein space of states and the unitarity issue with $J$-inner product somehow failed to draw his attention. While Toshiaki Tanaka treated null states as physical states in his study, the author regard them as unphysical since they correspond to complex eigenvalues. Toshiaki Tanaka concluded that $PT$-symmetric quantum mechanics is a different quantisation scheme rather than a generalisation of the standard Hermitian quantum mechanics. The author learned through \cite{Tanaka} that B. Bagchi, C. Quesne, M. Znojil studied the conservation of particle number in PT-symmetric quantum 
mechanics in \cite{Bagchi-Quesne-Znojil}. G. S. 
Japaridze also studied the conservation of particle number in PT-symmetric quantum mechanics in \cite{Japaridze}. In \cite{Everitt-Khalil-Zagoskin}, Mark J. Everitt at al. argue, without much detail, 
that $PT$-symmetric quantum mechanics is not a generalisation of the standard Hermitian quantum mechanics but merely another Hermitian quantum mechanics with a different inner product.
\appendix
\section{Real Structure and Fundamental Symmetry}
Let $V$ be the complex vector space of states of a quantum mechanical system. Define a map $J: V\longrightarrow \bar V$ by
\begin{equation}
\label{eq:real}
J\psi=\bar\psi
\end{equation}
for every $\psi\in V$, where $\bar V=\{\bar\psi:\psi\in V\}$ is the complex conjugate vector space. Then $J$ is a linear involution and is called a \emph{real structure} on $V$. Any $\psi\in V$ may be written as
$$\psi=\psi^++\psi^-$$
where
\begin{equation}
\begin{aligned}
 \psi^+&:=\frac{1}{2}(\psi+J\psi)=\frac{1}{2}(\psi+\bar\psi)=\psi_{\mathrm{re}},\\
 \psi^-&:=\frac{1}{2}(\psi-J\psi)=\frac{1}{2}(\psi-\bar\psi)=i\psi_{\mathrm{im}}.
 \end{aligned}
\end{equation}
We have
\begin{align*}
 J\psi^+&=\psi^+,\\
 J\psi^-&=-\psi^-.
\end{align*}
Hence one gets a direct sum of vector spaces
$$V=V^+\oplus V^-,$$
where
\begin{equation}
\begin{aligned}
 V^+&=\{\psi\in V:J\psi=\psi\},\\
 V^-&=\{\psi\in V:J\psi=-\psi\}.
\end{aligned}
 \end{equation}
Both $V^+$ and $V^-$ are real vector spaces. $V^+$ is isomorphic to $V^-$ via the linear map $\psi\longmapsto i\psi$. So,
$$\dim_{\mathbb{R}}V^+=\dim_{\mathbb{R}}V^-=\dim_{\mathbb{C}}V$$
if $\dim_{\mathbb{C}}V<\infty$. Let $V^+=V_{\mathbb{R}}$. Then $V^-$ can be written as $iV_{\mathbb{R}}$, so $V$ may be viewed as the complexification of $V_{\mathbb{R}}$
$$V_{\mathbb{R}}^{\mathbb{C}}=V_{\mathbb{R}}\otimes_{\mathbb{R}}\mathbb{C}.$$
The real structure $J$ in \eqref{eq:real} may be used to define an inner product $\langle\ ,\ \rangle$ on $V$ as follows: For any $\varphi,\psi\in V$,
\begin{equation}
\label{eq:hermitianproduct}
\begin{aligned}
 \langle\varphi,\psi\rangle&=\int_{-\infty}^\infty (J\varphi)\psi dx\\
 &=\int_{-\infty}^\infty\bar\varphi\psi dx.
 \end{aligned}
\end{equation}
\eqref{eq:hermitianproduct} is of course the standard Hermitian product known as Dirac braket $\langle\ |\ \rangle$ to physicists. Therefore, the real structure \eqref{eq:real} gives rise to the Hilbert space structure on $V$ as in the standard Hermitian quantum mechanics. In the paper, the author defined an inner product \eqref{eq:innerproduct} in a fashion that is more familiar to physicists. But from mathematical point of view, \eqref{eq:hermitianproduct} is more suitable form.

In studying a quantum mechanical system, it is not necessary to use complex-valued state functions but one may instead use real-valued state functions. This time, let $V$ be the real vector space of states of a quantum mechanical system. Any real-valued function may be written as the sum of an even function and an odd function. For any $\varphi(x)\in V$, $\varphi(x)$ can be written as
$$\varphi(x)=\frac{1}{2}[\varphi(x)+\varphi(-x)]+\frac{1}{2}[\varphi(x)-\varphi(-x)].$$
$\frac{1}{2}[\varphi(x)+\varphi(-x)]$ is an even function and $\frac{1}{2}[\varphi(x)-\varphi(-x)]$ is an odd function. Due to physical boundary conditions, these even and odd functions may be regarded as periodic functions. It is well-known that the Fourier series of a periodic even function includes only cosine terms, while the Fourier series of an odd function includes only sine terms. Define a map $J:V\longrightarrow V$ by
$$J\varphi(x)=\varphi(-x)$$
for every $\varphi(x)\in V$, i.e. $J=P$, the parity. Then $J$ is a linear involution. Any $\varphi(x)\in V$ may be written as
$$\varphi(x)=\varphi^+(x)+\varphi^-(x),$$
where
\begin{align*}
 \varphi^+(x)&=\frac{1}{2}[\varphi(x)+J\varphi(x)]=\frac{1}{2}[\varphi(x)+\varphi(-x)],\\
 \varphi^-(x)&=\frac{1}{2}[\varphi(x)-J\varphi(x)]=\frac{1}{2}[\varphi(x)-\varphi(-x)].
\end{align*}
We have
\begin{align*}
 J\varphi^+(x)&=\varphi^+(x),\\
 J\varphi^-(x)&=-\varphi^-(x).
\end{align*}
The map $J$ is called the \emph{fundamental symmetry}. One gets a direct sum of vector spaces
$$V=V^+\oplus V^-,$$
where
\begin{align*}
 V^+&=\{\varphi\in V: J\varphi=\varphi\},\\
 V^-&=\{\varphi\in V: J\varphi=-\varphi\}.
\end{align*}
$V^+$ is isomorphic to $V^-$ via the linear map $\varphi(x)\longmapsto\frac{d\varphi(x)}{dx}$. The fundamental symmetry may be used to define an inner product $\langle\ ,\ \rangle$ on $V$ as follows: For any $\varphi(x),\psi(x)\in V$,
\begin{equation}
\label{eq:ptproduct}
\begin{aligned}
 \langle\varphi(x),\psi(x)\rangle&=\int_{-\infty}^\infty (J\varphi(x))\psi(x) dx\\
 &=\int_{-\infty}^\infty\varphi(-x)\psi(x)dx.
\end{aligned}
\end{equation}
For any $\varphi^+(x)\in V^+$, $\varphi^-(x)\in V^-$, we have
\begin{align*}
\langle\varphi^+(x),\varphi^+(x)\rangle&=\int_{-\infty}^\infty(\varphi^+(x))^2dx\geq 0,\\
\langle\varphi^-(x),\varphi^-(x)\rangle&=-\int_{-\infty}^\infty(\varphi^-(x))^2dx\leq 0.
\end{align*}
The product of two even functions is an even function and the product of two odd functions is an even function. Thus, $\langle\ ,\ \rangle$ defines a positive definite and a negative definite inner products on $V^+$ and $V^-$, respectively. Moreover, for any $\varphi^+(x)\in V^+$, $\varphi^-(x)\in V^-$,
\begin{align*}
\langle\varphi^+(x),\varphi^-(x)\rangle&=\int_{-\infty}^\infty\varphi^+(x)\varphi^-(x)dx\\
 &=0,
\end{align*}
since the product of an even function and an odd function is always an odd function. Hence, $\varphi^+(x)$ and $\varphi^-(x)$ are orthogonal for all $\varphi^+(x)\in V^+,\varphi^-(x)\in V^-$, and so the direct sum $V^+\oplus V^-$ can be replaced by the orthogonal sum $V^+\dotplus V^-$. The fundamental symmetry gives rise to the Krein space structure on $V$. Although it may seem redundant, physically the inner product \eqref{eq:ptproduct} needs to be redefined as
\begin{equation}
\label{eq:ptproduct2}
 \begin{aligned}
 \langle\varphi(x),\psi(x)\rangle&=\int_{-\infty}^\infty \overline{(J\varphi(x))}\psi(x) dx\\
 &=\int_{-\infty}^\infty\bar\varphi(-x)\psi(x)dx
\end{aligned}
\end{equation}
for any $\varphi(x),\psi(x)\in V$. The reason is that in quantum mechanics $\langle\ ,\ \rangle$ often interacts with operators that may be complex.

\end{document}